\documentclass[
    ,final            
  ]
  {aipproc}

\layoutstyle{6x9}

\graphicspath{{Pics/}}
\DeclareGraphicsExtensions{.eps,.ps}

\def\muf{{\mu^{}_f}}
\def\mur{{\mu^{}_r}}

\def\alphas{{\alpha_s}}
\def\MSbar{{$\overline{\mbox{MS}}\,$}}

\begin{document}
\vspace*{-20mm}
DESY 11-113
\vspace*{10mm}

\title{Running heavy-quark masses in DIS}


\author{S.~Alekhin}{
  address={
Deutsches Elektronensynchrotron DESY, 
    Platanenallee 6, D--15738 Zeuthen, Germany
}
 ,altaddress={
    Institute for High Energy Physics, 
    142281 Protvino, Moscow region, Russia
}
}

\author{S.~Moch}{
 address={
Deutsches Elektronensynchrotron DESY, 
    Platanenallee 6, D--15738 Zeuthen, Germany
}
}

\begin{abstract}
We report on determinations of the running mass for charm quarks from deep-inelastic scattering reactions. 
The method provides complementary information on this fundamental parameter from hadronic processes with space-like kinematics.
The obtained values are consistent with but systematically lower than the world average as published by the PDG.
We also address the consequences of the running mass scheme for heavy-quark parton distributions 
in global fits to deep-inelastic scattering data.
\end{abstract}

\maketitle


Quark masses are fundamental parameters of the gauge theory of the strong interactions, Quantum Chromodynamics (QCD).
They are, however, not directly observable due to confinement.
Rather one has to compute the dependence of cross sections or other measurable quantities 
on the heavy quark mass including higher order radiative corrections and renormalization --
a procedure which requires the choice of a scheme for the definition the mass parameter.
There exists a variety of schemes for heavy-quark masses. 
The most popular ones are the on-shell and the \MSbar scheme.
In the former the so-called pole-mass $m$ coincides with the pole 
of the heavy-quark propagator at each order in perturbative QCD. 
This definition, however, has intrinsic theoretical limitations 
with ambiguities of order ${\cal O}(\Lambda_{\rm QCD})$ and a strong dependence 
of the value of the mass parameter on the order of perturbation theory.
The \MSbar scheme is one of the so-called short-distance masses for heavy quarks.
It realizes the concept of a running mass $m(\mu)$ depending on the scale $\mu$ 
of the hard scattering in complete analogy to the treatment 
of the running coupling $\alphas(\mu)$. 
As a benefit, predictions for hard scattering cross sections in terms of the \MSbar mass 
display better convergence properties and greater perturbative stability 
at higher orders.

Cross sections for the production of heavy-quarks in deep-inelastic scattering (DIS) 
are particularly well suited to confront the quark mass dependence
of theoretical predictions in perturbative QCD with experimental measurements in space-like kinematics.
For the production of charm quarks in neutral (NC) or charged current (CC) 
DIS there exists precise data from the HERA collider and fixed-target experiments. 
In QCD the DIS heavy-quark structure functions $F_k$ which parametrize the hadronic
cross section are subject to the standard factorization 
\begin{equation}
  \label{eq:totalF2c}
  F_k(x,Q^2,m^2) =
  \sum\limits_{i = q,{\bar{q}},g} \,\,
  \biggl[ f_{i}(\mu^2)\, \otimes 
  C_{k, i}\left(Q^2,m^2,\alphas(\mu^2)\right) \biggr](x)
  \, ,
  \qquad\qquad k = 1,2,3
  \, ,
\end{equation}
where the perturbative coefficient functions $C_{k, i}$ are known 
to next-to-leading order (NLO) for CC~\cite{Gottschalk:1980rv,Gluck:1996ve} 
and approximately to next-to-next-to-leading order (NNLO) for NC~\cite{Laenen:1992zk,Presti:2010pd}.
$Q^2$ and $x$ are the usual DIS kinematical variables and $m$ is the heavy-quark (pole) mass.
In eq.~(\ref{eq:totalF2c}) we also display all dependence on the other non-perturbative parameters, 
i.e. the parton distribution functions $f_i$ (PDFs) for light quarks and gluons 
as well as the strong coupling constant $\alphas$.
The running mass definition can be implemented in
eq.~(\ref{eq:totalF2c}) to the respective order in perturbation theory 
by simply following the standard procedure for changing the renormalization condition, 
i.e. $m \to m(\mu)$ see ref.~\cite{Alekhin:2010sv}.

The present study has two aspects:
First of all we determine the \MSbar charm mass $m_c(m_c)$
from DIS in a variant of the global analysis~\cite{Alekhin:2009ni} and 
compare to the world average as published by the PDG.
Secondly, we investigate the improvements in the uncertainty of heavy-quark PDFs 
in a global fit within the fixed-flavor number scheme (FFNS) when the running mass scheme is applied.

\bigskip
\noindent
{\bf 1.\,\, Running charm-quark mass}\\
\indent
The parametric dependence of the DIS structure functions $F_k$ in eq.~(\ref{eq:totalF2c}) 
on $m$ can be used for a determination of the heavy-quark mass. 
The sensitivity of this procedure relates directly to the corresponding uncertainty on the measurements of $F_k$.
E.g., for charm production in NC DIS the nucleon structure function $F_2$ yields 
$\Delta m_c / m_c \simeq 0.75 \, \Delta F_2 / F_2\, ,$
which implies that an experimental accuracy of 8\% for $F_2$ 
translates into an uncertainty of 6\% for the charm-quark mass~\cite{Alekhin:2010sv}.
With the precision of current DIS data for charm production 
this suggests an error on $m_c(m_c)$ of ${\cal O}({\rm few}) \%$ 
as the ultimate precision in the approach based on inclusive structure functions.

Starting from eq.~(\ref{eq:totalF2c}) we have extracted the \MSbar charm mass $m_c(m_c)$ 
in a phenomenological study similar to~\cite{Alekhin:2008hc}, 
i.e. a global fit including fixed-target (CCFR~\cite{Bazarko:1994tt}, NuTeV~\cite{Goncharov:2001qe})
and collider data~\cite{Chekanov:2003rb,Aaron:2009ut} in the FFNS 
(with $n_f = 3$) as a variant of the ABKM one~\cite{Alekhin:2009ni}.
We have obtained~\cite{Alekhin:2010sv},
\begin{eqnarray}
  \label{eq:mbarcnlo}
  m_c(m_c) &=& 1.26\,\, \pm 0.09 ({\rm exp})\,\, \pm 0.11 ({\rm th})\,\,
  {\rm GeV}\qquad\qquad
  {\rm at\,\, NLO}
  \, ,
  \\[2ex]
  \label{eq:mbarcnnlo}
  m_c(m_c) &=& 1.01\,\, \pm 0.09 ({\rm exp})\,\, \pm 0.03 ({\rm th})\,\,
  {\rm GeV}\qquad\qquad
  {\rm at\,\, NNLO}_{\rm approx}
  \, ,
\end{eqnarray}
to NLO and approximate NNLO in perturbation theory.
The quoted experimental uncertainty results from the propagation of the statistical and
systematic errors in the data with account of error correlations whenever available.
The theoretical uncertainty is estimated from the variation of the renormalization and factorization scales for the choice 
$\mur^2 = \muf^2 = Q^2 + \kappa m_c^2$ for $F_2$ with $\kappa$ in the range $\kappa \in [0,8]$.
For consistency it has been checked that different scale choices 
do not deteriorate the statistical quality of the fit. 
Besides the charm mass $m_c$ quoted in eqs.~(\ref{eq:mbarcnlo}) and (\ref{eq:mbarcnnlo}) 
and the value of the strong coupling constant determined to
$\alphas(M_Z) = 0.1135 \pm 0.0014$ at NNLO~\cite{Alekhin:2009ni}, 
the global fit relies on the same 22 PDF parameters as in ABKM~\cite{Alekhin:2009ni}.
This is particularly important in order to control simultaneously the sensitivity to the strange-quark PDF 
in CC case which is correlated with $m_c$ due to the Born process $W^\pm s \to c$. 

At this conference, NOMAD has reported a new analysis of CC fixed-target data.
The increased precision due to the very high statistics has been used to 
determine the running mass $m_c(m_c)$ to NLO exclusively from CC DIS 
leading to the (preliminary) result~\cite{Petti:2011} 
\begin{eqnarray}
  \label{eq:mbarcnloCC}
  m_c(m_c) &=& 1.070\,\, \pm 0.067 ({\rm exp})\,\, \pm 0.050 ({\rm th})\,\,
  {\rm GeV}\qquad\qquad
  {\rm at\,\, NLO\,\,}
  \, .
\end{eqnarray}

One has to compare the numbers in eqs.~(\ref{eq:mbarcnlo}), (\ref{eq:mbarcnnlo}) and (\ref{eq:mbarcnloCC}) with the 
world average of the PDG~\cite{Nakamura:2010pdg} quoted in the \MSbar scheme 
as $m_c(m_c) = 1.27\,\,^{+0.07}_{-0.09}\,\, {\rm GeV} \, ,$
which is entirely based on lattice computations 
or analyses of experimental data with time-like kinematics 
from $e^+e^-$-collisions, e.g. with the help of QCD sum rules.
It is therefore interesting to note that the DIS results in 
eqs.~(\ref{eq:mbarcnlo}), (\ref{eq:mbarcnnlo}) and (\ref{eq:mbarcnloCC}) 
for hadronic processes with space-like kinematics 
are consistent with but systematically lower than the PDG value. 
In order to understand at least one source of this deviation, one should note 
the QCD sum rules analyses typically assume the Bethke world average 
for the value of the strong coupling constant~\cite{Bethke:2009jm}, which is 
$\alphas(M_Z) = 0.1184 \pm 0.0007$ 
and therefore significantly larger than the ABKM result,
see also~\cite{Alekhin:2011gj} and references therein.
A recent determination from a charmonium QCD sum rules analysis~\cite{Dehnadi:2011gc} 
quotes a central value $m_c(m_c) = 1.257 \pm 0.026~$GeV using the world average 
and parametrizes separately the dependence of $m_c(m_c)$ on value of $\alphas(M_Z)$. 
Using the ABKM value $\alphas(M_Z) = 0.1135$ instead one extracts from Table~15 in \cite{Dehnadi:2011gc} 
a value of $m_c(m_c) = 1.257 \pm 0.026~$GeV, i.e. 
a systematic shift downwards at the level on $1 \sigma$ statistical uncertainty 
bringing QCD sum rule analysis from $e^+e^-$-collisions in better agreement with the 
DIS results in eqs.~(\ref{eq:mbarcnlo}), (\ref{eq:mbarcnnlo}) and (\ref{eq:mbarcnloCC}).
Note that the latter determinations account for the full correlation 
of the dependence on $m_c$ and $\alphas$ through a simultaneous fit of the parameters.

\begin{figure}[ht]
\hspace*{-10mm}
  \includegraphics[width=0.675\columnwidth,height=9.5cm]{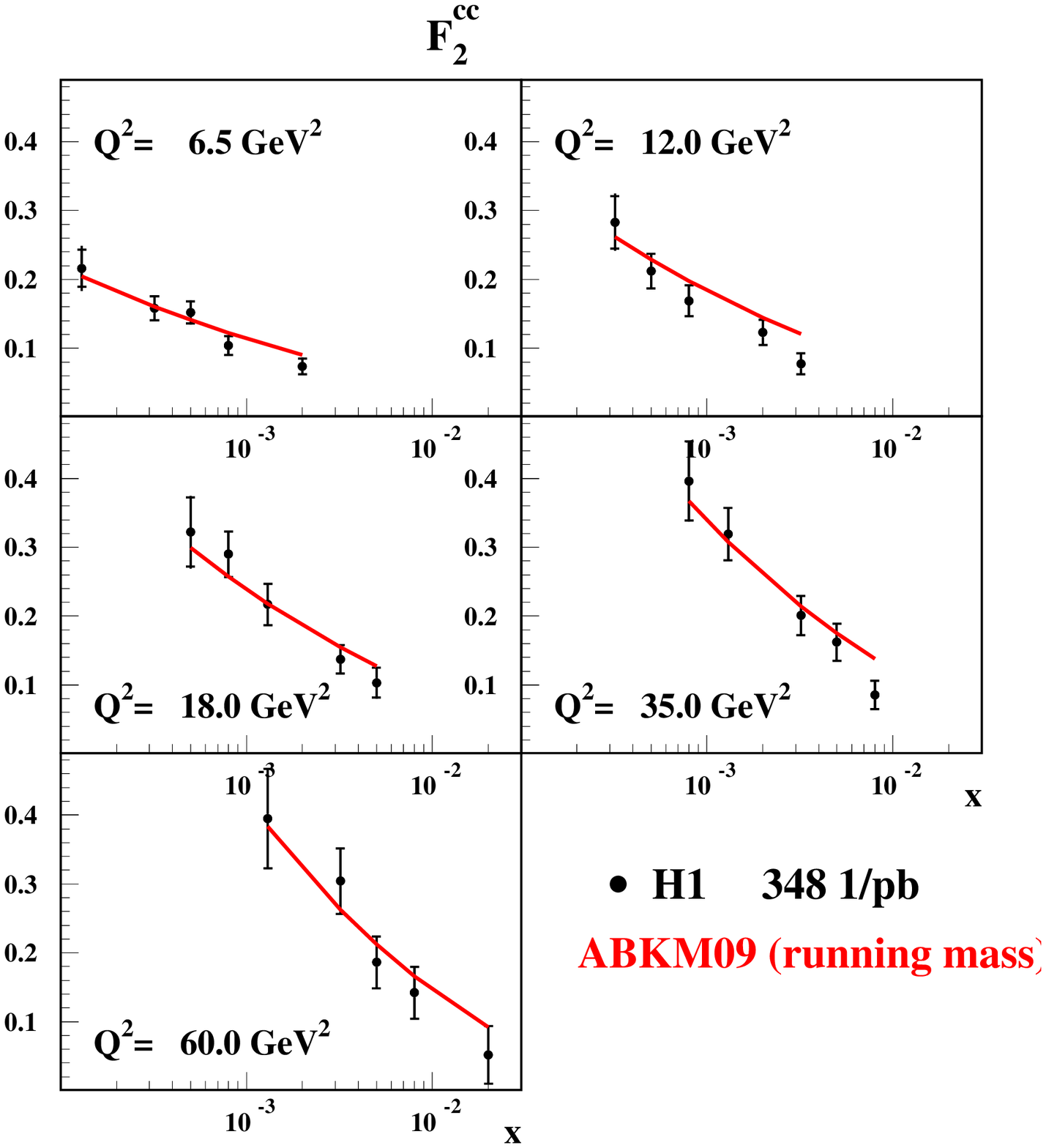}
\hspace*{-5mm}
  \includegraphics[width=0.45\columnwidth,height=9.25cm]{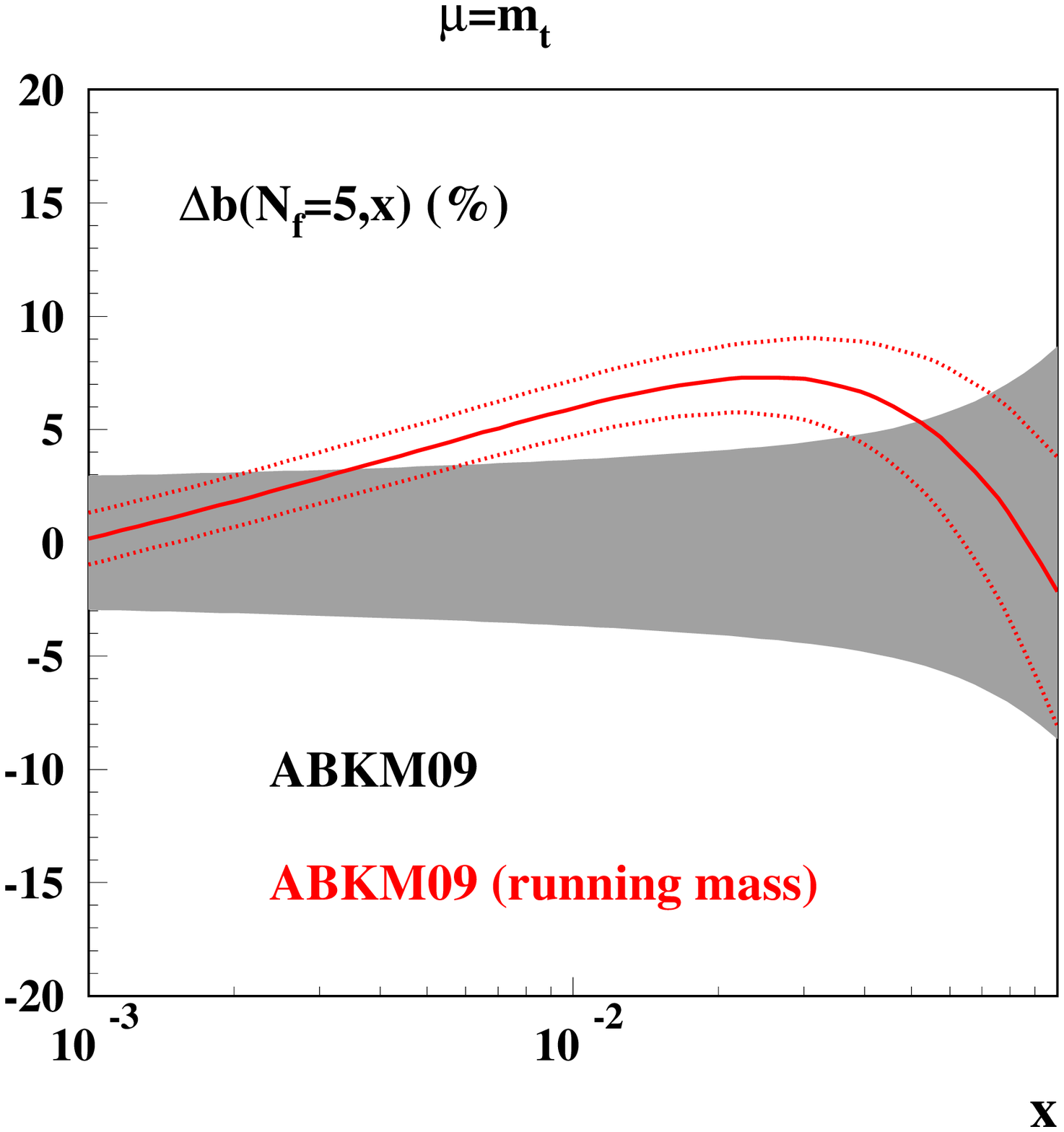}
  \caption{
    \label{fig:hq-pdf}
Left:
      The data on the charm structure function $F_2^{c{\bar c}}$ 
      by the H1 collaboration~\cite{Aaron:2011gp} 
      in comparison with the running-mass scheme predictions of ref.~\cite{Alekhin:2010sv}.
Right:
      The $b$-quark PDF uncertainties obtained in the global fits:
      The dotted (red) lines denote the $\pm 1 \sigma$ band of relative
      uncertainties (in percent) and the solid (red) line indicates the
      difference in the central prediction resulting from the change of the
      mass scheme and using $m_b(m_b) = 4.19~$GeV (bottom).
      For comparison the shaded (grey) area represents the uncertainties in
      the ABKM fit~\cite{Alekhin:2009ni}.
}
\end{figure}

Including the PDG constraint on $m_c(m_c)$ into the fit of
ref.~\cite{Alekhin:2010sv} we get the value of $m_c(m_c) = 1.18 \pm 0.06~$GeV at NNLO.
The recent data on the charm structure function $F_2^{c{\bar c}}$ at low $Q^2$ published 
by the H1 collaboration~\cite{Aaron:2011gp}, which were not included into the fit of ref.~\cite{Alekhin:2010sv},
are compared with the running-mass scheme predictions of ref.~\cite{Alekhin:2010sv} obtained with this value of $m_c(m_c)$ in fig.~\ref{fig:hq-pdf} (left). 
The agreement is quite good therefore these data also prefer somewhat smaller value of $m_c(m_c)$ than the PDG average.

\bigskip
\noindent
{\bf 2.\,\, Heavy quark PDFs}\\
\indent
At asymptotically large scales $Q \gg m_c, m_b$ 
the genuine heavy quark contributions in a FFNS with $n_f = 3$ grow as $\alpha_s(Q^2) \ln(Q^2/m^2)$ 
and can be resummed by means of standard renormalization group methods. 
This procedure leads to so-called heavy quark PDFs 
in theories with effectively $n_f = 4$ and $n_f = 5$ light flavors, which are
the appropriate descriptions for processes at LHC energies.
The PDFs for charm- and bottom-quarks in $n_f=4$- and $n_f=5$-flavor schemes 
are generated from the light flavor PDFs in a $n_f=3$-flavor FFNS 
as convolutions with massive operator matrix elements (OMEs), see e.g.~\cite{Alekhin:2009ni}.
The uncertainty on heavy-quark PDFs is therefore directly related 
to the accuracy of the quark masses $m_c$ or $m_b$, which appear 
parametrically in the OMEs.
This uncertainty can be significantly reduced 
through the use of the \MSbar scheme, which, of course, has to be applied also to the massive OMEs.

In our global fit~\cite{Alekhin:2010sv} $m_c(m_c)$ has been left a free parameter 
supplemented by the PDG constraint. 
In this way we have generated a charm-PDF with comparable uncertainties to the
one of~\cite{Alekhin:2009ni} (which has used the pole mass definition for $m_c$).
For $m_b$ the currently available DIS data displays no sensitivity at all and 
we have constrained $m_b(m_b)$ directly to its PDG
value~\cite{Nakamura:2010pdg}, i.e. $m_b(m_b) = 4.19\,\,^{+0.18}_{-0.06}~${\rm GeV}.
As shown in fig.~\ref{fig:hq-pdf} (right) the uncertainty of the resulting bottom-PDF is greatly reduced.
This improvement will certainly have an impact on LHC phenomenology, 
e.g. allowing for precise predictions for the production of single-top-quarks.

\bibliographystyle{aipproc}   

\bibliography{dis2011}

\IfFileExists{\jobname.bbl}{}
 {\typeout{}
  \typeout{******************************************}
  \typeout{** Please run "bibtex \jobname" to optain}
  \typeout{** the bibliography and then re-run LaTeX}
  \typeout{** twice to fix the references!}
  \typeout{******************************************}
  \typeout{}
 }

\end{document}